\documentstyle[11pt,aaspp4]{article}

\def\etal {\emph{et~al.\,}}

\def\go {\mathrel{\raise.3ex\hbox{$>$}\mkern-14mu\lower0.6ex\hbox{$\sim$}}}
\def\lo {\mathrel{\raise.3ex\hbox{$<$}\mkern-14mu\lower0.6ex\hbox{$\sim$}}}

\begin{document}

\title{Monte-Carlo Simulations of Globular Cluster Evolution. \\
I. Method and Test Calculations.}

\author{Kriten J.~Joshi\altaffilmark{1}, Frederic 
A.~Rasio\altaffilmark{2,3}, and Simon Portegies Zwart\altaffilmark{4,5}}

\affil{Department of Physics, Massachusetts Institute of Technology}

\altaffiltext{1}{6-218M MIT, 77 Massachusetts Ave, Cambridge, MA 02139; 
email: kjoshi@mit.edu.}
\altaffiltext{2}{6-201 MIT, 77 Massachusetts Ave, Cambridge, MA 02139; 
email: rasio@mit.edu.}
\altaffiltext{3}{Alfred P.\ Sloan Research Fellow.}
\altaffiltext{4}{Boston University, 725 Commonwealth Ave, Boston, MA 02215; 
email: spz@komodo.bu.edu.}
\altaffiltext{5}{Hubble Fellow.}

\begin{abstract}

We present a new parallel supercomputer implementation of the Monte-Carlo
method for simulating the dynamical evolution of globular star clusters.
Our method is based on a modified version of H\'enon's Monte-Carlo algorithm 
for solving the Fokker-Planck equation. Our code allows us to follow
the evolution of a cluster containing up to $5\times10^5$ stars to core 
collapse in $\lo 40$ hours of computing time.
In this paper we present the results of test calculations for 
clusters with equal-mass stars, starting from both Plummer and 
King model initial conditions. We consider isolated as well as 
tidally truncated clusters.
Our results are compared to those obtained from approximate, self-similar 
analytic solutions, from direct numerical integrations of the 
Fokker-Planck equation, and from direct $N$-body integrations 
performed on a GRAPE-4 special-purpose computer with $N=16384$. 
In all cases we find excellent agreement with other methods,
establishing our new code as a robust tool for the numerical study of
globular cluster dynamics using a realistic number of stars.
\end{abstract}


\section{Introduction}

The dynamical evolution of dense star clusters is a problem of fundamental
importance in theoretical astrophysics, but many aspects of the problem have
remained unresolved in spite of years of numerical work and improved
observational data. On the theoretical side, some key unresolved issues
include the role played by primordial binaries and their dynamical interactions
in the overall cluster dynamics and in the production of exotic sources
(Hut \etal 1992), and the importance of tidal shocking for the
long-term evolution and survival of globular clusters in the Galaxy 
(Gnedin, Lee \& Ostriker 1999).
On the observational side, we now have many large data sets providing a
wealth of information on blue stragglers, X-ray sources and millisecond
pulsars, all found in large numbers in dense clusters 
(e.g., Bailyn 1995; Camilo \etal 2000; Piotto \etal 1999).
Although it is clear that these objects are produced at high rates 
through dynamical interactions in the dense cluster cores, the details of 
the formation mechanisms, and in particular the interplay between binary 
stellar evolution and dynamical interactions, are far from understood.

\subsection{Overview of Numerical Methods}

Following the pioneering work of H\'enon (1971a,b), many numerical simulations 
of globular cluster evolution
were undertaken in the early 1970's, by two groups, at Princeton and Cornell,
using different Monte-Carlo methods, now known as the ``Princeton method'' and
the ``Cornell method'' (see Spitzer 1987 for an overview of the methods). 
In the Princeton method, the orbit of each star is integrated
numerically, while the diffusion coefficients for the change in velocity 
$\Delta \mathbf{v}$ and $(\Delta v)^2$ (which are calculated analytically)
are selected to represent the average 
perturbation over an entire orbit. Energy conservation is enforced by requiring
that the total energy be conserved in each radial region of the cluster.
The Princeton method assumes an isotropic, Maxwellian velocity distribution 
of stars to compute the diffusion coefficients, and hence does not take in to
account the anisotropy in the orbits of the field stars. One advantage of this
method is that, since it follows the evolution of the cluster on a dynamical timescale,
it is possible to follow the initial ``violent relaxation'' phase more easily.
Unfortunately, for the same reason, it also requires considerably more computing
time compared to other versions of the Monte-Carlo method.  
In the Cornell method, also known as the 
``Orbit-averaged Monte-Carlo method'', the changes in energy $E$ and angular
momentum $J$ per unit time (averaged over an orbit) are computed analytically
for each star.
Hence, the time consuming dynamical integration of the orbits is not required.
In addition, since the diffusion coefficients are computed for both $\Delta E$
\emph{and} $\Delta J$, the Cornell method does take in to account the anisotropy
in the orbits of the stars. The ``H\'enon method''
is a variation of the Cornell method, in which the velocity perturbations
are computed by considering an encounter between pairs of neighboring stars.
This also allows the local 2-D phase space distribution $f(E,J)$ to
be sampled correctly. Our code is based on a modified version of H\'enon's 
method. We have modified H\'enon's algorithm for determining the timestep 
and computing the representative encounter between neighboring stars. 
Our method allows the timestep to be made much smaller in order to resolve 
the dynamics in the core more accurately. We describe the basic method
and our modifications in more detail below in \S 2.

The Monte-Carlo methods were first used to study the development of the
gravothermal instability (Spitzer \& Hart 1971a,b; H\'enon 1971a,b) and to 
explore the effects of a massive black hole at the center of a globular cluster
(Lightman \& Shapiro 1977). In those early studies, the available computational 
resources limited the number of particles
used in the Monte-Carlo simulations to $\lo 10^3$. Since this is much smaller 
than the real number of stars in a globular cluster ($N\sim10^5-10^6$),  
each particle in the simulation represents effectively a whole spherical 
shell containing many stars, and the method provides no information about 
individual objects and their dynamical interactions.
More recent implementations have used up to $\sim 10^4-10^5$ particles and 
have established the method as a promising alternative
to direct $N$-body integrations (Stod\'olkiewicz 1986; Giersz 1998).
Monte-Carlo simulations have also been used to study specific
interaction processes in globular clusters, such as tidal capture 
(Di Stefano \& Rappaport 1994), interactions involving primordial
binaries (Hut, McMillan, \& Romani 1992) and stellar evolution 
(Portegies Zwart \etal 1997).
However, in all these studies the background cluster was assumed to have
a {\em fixed structure\/}, which is clearly not realistic.
The main goal of our study is to perform Monte-Carlo simulations
of cluster dynamics treating both the cluster itself and all relevant
interactions self-consistently, including
all dynamical interactions involving primordial binaries.
This idea is particularly timely because the latest generation of
parallel supercomputers now makes it possible
to do such simulations for a number of objects equal to the actual number of
stars in a globular cluster.  
Using the correct number of stars in a cluster simulation
ensures that the relative rates of different dynamical processes (which all
scale differently with the number of stars) are correct. This is
crucial if many different dynamical processes are to be 
incorporated, as we plan to do in this study.

In addition to Monte-Carlo and $N$-body simulations, a new method
was developed, mainly by Cohn and collaborators, based on the direct numerical 
integration of the orbit-averaged Fokker-Planck equation
(Cohn 1979, 1980; Statler, Ostriker \& Cohn 1987; Murphy \& Cohn 1988).
Unlike the Monte-Carlo methods, the direct Fokker-Planck method constructs 
the (smooth) distribution function of the system
on a grid in phase space, effectively providing the $N\to\infty$ limit
of the dynamical behavior. The original formulation of the method used 
a 2-D phase space distribution function $f(E,J)$ (Cohn 1979). 
However, the method was later reduced to a 1-D form using an isotropized 
distribution function $f(E)$ (Cohn 1980). 
The reduction of the method to one dimension speeded up the 
calculations significantly. In addition, the use of the Chang \& Cooper (1970)
differencing scheme provided much better energy conservation compared 
to the original 2-D method. The 1-D method provided very good results 
for isolated clusters, in which the effects of velocity anisotropy are 
small. The theoretically predicted emergence of a power-law density 
profile in the late stages of evolution for isolated
single-component systems has been clearly verified using
this method (Cohn 1980). Calculations that include the effects of binary
interactions, including primordial binaries, have also allowed the
evolution to be followed beyond core collapse (Gao \etal 1991).
However, results obtained using the 1-D method showed 
substantial disagreement with $N$-body results for tidally truncated 
clusters, in which the evaporation rate is dramatically affected by the 
velocity anisotropy. Ignoring the velocity anisotropy led to a significant 
overestimate of the evaporation rate from the cluster, resulting in shorter 
core-collapse times for tidally truncated clusters 
(Portegies Zwart \etal 1998).
A recent implementation of the Fokker-Planck method by Drukier \etal 
(1999) has extended the algorithm to allow a 2-D distribution 
function, while also improving the energy conservation. 
A similar 2-D method has also been 
developed by Takahashi (1995, 1996, 1997). The new implementations 
produce much better agreement with $N$-body results (Takahashi
\& Portegies Zwart 1998), and can also model the effects of mass loss due
to stellar evolution (Takahashi \& Portegies Zwart 1999), as well as
binary interactions (Drukier \etal 1999).

For many years direct $N$-body simulations were limited to systems with 
$N \lo 10^3$ stars. New, special-purpose computing hardware such as the 
GRAPE (Makino \etal 1997) now make it possible to perform
direct $N$-body simulations with up to $N\sim10^5$ single stars 
(Hut \& Makino 1999), but the inclusion of a significant fraction
of primordial binaries in these simulations remains prohibitively expensive.
The large dynamic range of the orbital timescales of the stars in the
cluster presents a serious difficulty for $N$-body simulations. 
The orbital timescales can be as small as the periods of the tightest
binaries. The direct integration of stellar orbits is especially plagued 
by this effect. These difficulties are overcome using techniques
such as individual integration timesteps, and various schemes for
regularizing binaries (see, e.g., Aarseth 1998 for a
review). These short-cuts introduce specific selection effects,
and complicate code development considerably.
Instead, in the Monte-Carlo methods, individual stellar orbits are represented
by their constants of the motion (energy $E$ and angular momentum $J$ for
a spherical system) and perturbations to these orbits are computed periodically
on a timestep that is a fraction of the relaxation time. Thus the numerical
integration proceeds on the natural timescale for the overall dynamical
evolution of the cluster. Note also that, because of exponentially growing 
errors in the direct integration
of orbits, $N$-body simulations, just like Monte-Carlo simulations,
can only provide a statistically correct representation of cluster dynamics
(Goodman \etal 1993; Hernquist, Hut, \& Makino 1993).

A great advantage of the Monte-Carlo method is that it makes it particularly
easy to add more complexity and realism to the simulations one layer at a time.
The most important processes that we will focus on initially will be stellar
evolution and mass loss through a tidal boundary.
Interactions of single stars with primordial binaries, binary-binary 
interactions, stellar evolution in binaries, and a detailed 
treatment of the influence of the Galaxy, including tidal shocking of the 
cluster when it passes through the galactic disk, will be incorporated
subsequently. 

Recent improvements in algorithms and available computational resources
have allowed meaningful comparisons between the
results obtained using different numerical methods 
(see for example the ``Collaborative Experiment'' by Heggie \etal 1999).
However, there still remain substantial unresolved differences between the 
results obtained using various methods. For example, the lifetimes of clusters 
computed
recently using different methods have been found to vary significantly.
Lifetimes of some clusters computed using direct Fokker-Planck simulations by
Chernoff \& Weinberg (1990) are up to an order of magnitude shorter than those
computed using $N$-body simulations and a more recent version of the 
Fokker-Planck method (Takahashi \& Portegies Zwart 1998). 
It has been found that, in many cases, the
differences between the two methods can be attributed to the lack of an
appropriate discrete representation of the cluster in the Fokker-Planck 
simulations. This can lead to an over-estimate of the
mass-loss rate from the cluster, causing it to disrupt sooner.
Recently, new calibrations of the mass loss in the Fokker-Planck method
(Takahashi \& Portegies Zwart 1999) that account for the slower mass loss in
discrete systems, has led to better agreement between the methods. 
The limitation of $N$-body 
simulations to small $N$ (especially for clusters containing a large 
fraction of primordial binaries)
makes it particularly difficult to compare the results 
with Fokker-Planck calculations, which are effectively done for
very large $N$ (Portegies Zwart \etal 1998, Heggie \etal 1999). 
This gap can be filled very naturally with Monte-Carlo
simulations, which can be used to cover the entire range of $N$'s
not accessible by other methods.

\subsection{Astrophysical Motivation}

The realization over the last 10 years that primordial binaries are
present in globular clusters in dynamically significant numbers has completely
changed our theoretical perspective on these systems (see. e.g., the review by
Hut \etal 1992). Most importantly,
dynamical interactions between hard primordial binaries and other single
stars or binaries are now thought to be the primary mechanism for supporting
a globular cluster against core collapse (McMillan, Hut, \& Makino 1990, 1991;
Gao \etal 1991). 
In addition, exchange interactions between primordial binaries and 
compact objects can explain very naturally
the formation of large numbers of X-ray binaries and recycled pulsars in
globular cluster cores (Sigurdsson \& Phinney 1995; Davies \& Hansen 1998;
Portegies Zwart \etal 1997). 
Previously, it was thought that primordial binaries
were essentially nonexistent in globular clusters, and so other mechanisms
such as tidal capture and three-body
encounters had to be invoked in order to form binaries dynamically during
core collapse. However, these other mechanisms have some serious
problems, and are much more likely to result in mergers than in the
formation of long-lived binaries (Chernoff 1996; Kochanek 1992; Kumar
\& Goodman 1996). 

Hubble Space Telescope (HST) observations have provided direct constraints on 
primordial binary fractions in clusters. The binary fraction is a key 
input parameter for any realistic study of cluster dynamics.
For example, the recent observation of a broadened main sequence in NGC 6752,
based on HST PC images of its core, suggest that the binary
fraction is probably in the range 15\%--38\% in the inner core
(Rubenstein \& Bailyn 1997).

Despite the fact that binaries play a crucial role in the late phases
of evolution of a cluster, the overall
evolution of a binary population within a cluster, and its direct implications
for the formation rate of observable binaries and blue stragglers remains 
poorly understood.
In addition, the relative importance of binaries in a cluster, like many
other physical processes, may depend on the actual size ($N$) of the cluster.
This makes it difficult to extend results obtained from smaller $N$-body
simulations to realistic globular cluster models.
When the initial primordial binary fraction is below a certain critical value,
a globular cluster core can run out of binaries before the end of its lifetime,
i.e., before being evaporated in the tidal field of the Galaxy (McMillan \&
Hut 1994). Without the support of binaries, the cluster will undergo a
much deeper core collapse and so-called gravothermal oscillations
(Sugimoto \& Bettwieser 1983; Breeden \etal 1994; Makino 1996). 
At maximum contraction,
the core density may increase by many orders of magnitude, leading to
greatly enhanced interaction rates.
Our new Monte-Carlo code  will allow us to follow the evolution of a cluster
through this phase, including in detail the dynamical interactions between
the $\sim 10^3$ objects in the core. 

Of particular interest is the possibility
that successive collisions and mergers of MS stars might lead to a {\em 
runaway process\/}.
The recent HST observations of stellar cusps in the cores of M15
(Guhathakurta \etal 1996, Sosin \& King 1997) and NGC 6624 (Sosin \& King
1995) have generated renewed interest in the possibility of massive black 
holes in globular clusters. The most significant unresolved theoretical issue 
concerns the manner in which such a black hole could form in a dense cluster. 
One of the likely routes, which we plan to examine with our simulations, 
is via the collisions and mergers of main-sequence stars, leading to the 
runaway build-up of a massive object and its eventual gravitational 
collapse (Portegies Zwart \etal 1999).

A very significant effect of the galactic environment on a cluster is the
gravitational shock heating of the cluster due to passages close to the bulge
and through the disk. When a cluster passes through the Galactic disk, it
experiences a time-varying gravitational force that pulls the cluster
toward the equatorial plane. The net effect of the shock is to induce an
increase in the average energy of the stars, causing the binding energy
of the cluster to decrease, and the rate of escape of stars through evaporation
to increase (Chernoff, Kochanek, \& Shapiro 1986). In addition, 
in some cases, ``shock-induced relaxation'' can be almost as important 
as two-body relaxation in the overall evolution of the cluster 
(Gnedin, Lee \& Ostriker 1999; Gnedin \& Ostriker 1997). 
Both the energy shift and the relaxation induced by tidal 
shocking can be incorporated in our Monte-Carlo method by assuming an orbit 
for the cluster around the Galactic center and introducing an appropriate
perturbation to the energy of the stars each time the cluster passes
through the disk. This can be done without adding much computational 
overhead to the problem, since tidal shocking only occurs twice during 
the orbital period of the cluster. The ability of the Monte-Carlo method 
to model such effects simultaneously with a realistic treatment of the 
internal dynamical evolution of the cluster makes it a very useful tool 
in verifying and extending previous results obtained using other methods.

The star-by-star representation of the system in Monte-Carlo simulations 
makes it easy to study of the evolution of a particular population of stars within 
a cluster. For example, the evolution of a population of neutron stars
could be followed closely, to help predict their properties and 
expected distributions within clusters. Of particular interest are
M15 and 47 Tuc, which have both been the targets of 
several highly successful searches for pulsars (Anderson 1992;
Robinson \etal 1995; Camilo \etal 2000).
The observed properties of pulsars in these clusters are found to
be very different. The pulsars in 47 Tuc are all millisecond pulsars,
and most are in short-period binaries, while those in M15 are mostly 
single recycled pulsars with longer pulse periods. This suggests that 
these two clusters may provide very different dynamical environments 
for the formation of recycled pulsars.

\section{The Monte-Carlo Method}

\subsection{Overview}

Our basic algorithm for doing stellar dynamics is based on the 
``orbit-averaged Monte-Carlo method'' developed by H\'enon (1971a,b). 
The method was later used and improved by Stod\'olkiewicz (1982, 1985, 1986). 
It has also recently been used by Spurzem \& Giersz (1996) to follow the 
evolution of hard three-body binaries in a cluster with equal point-mass 
stars. New results using Stod\'olkiewicz's version of the method were also
presented recently by Giersz (1998). In earlier implementations of 
the Monte-Carlo method with $N \sim 10^3$, each particle in the simulation 
was a ``superstar,'' representing many individual stars
with similar orbital properties. In our implementation, with 
$N \sim 10^5 - 10^6$, we treat each particle in the simulation as a 
single star. We have also modified H\'enon's original algorithm to 
allow the timestep to be made much smaller in order to resolve the 
dynamics in the core more accurately.

In the simplest case of a spherical system containing
$N$ point masses the algorithm can be summarized as follows.
We begin by assigning to each star a mass, radius and velocity by sampling
from a spherical and isotropic distribution function (for example,
the Plummer model).
Once the positions and masses of all stars are known, the gravitational
potential of the cluster is computed assuming spherical symmetry.
The energy and angular momentum of each star are then calculated. Energy and
angular momentum are perturbed at each timestep to simulate the effects of 
two-body and three-body relaxation. The perturbations depend
on each star's position and velocity, and on the density of stars in its
neighborhood. The timestep should be a fraction of the relaxation time
for the cluster (which is larger than the dynamical time
by a factor $\propto N/\ln N$). The perturbation of the energy and angular
momentum of a star at each timestep therefore represents the cumulative
effect of many small (and distant) encounters with other stars.
Under the assumption of spherical symmetry, the cross-sections
for these perturbations can be computed analytically.
The local number density is computed using a sampling procedure. Once
a new energy and angular momentum is assigned to each star, a new
realization of the system is generated by assigning to each star a new
position and velocity in an orbit that is consistent with its new energy
and angular momentum. In selecting a new position for each star along its
orbit, each position is weighted by the amount of time the star spends
around that position. Using the new positions, the gravitational potential
is then recomputed for the entire cluster. This procedure is then repeated
over many timesteps. After every timestep, all stars with positive total 
energy (cf. \S 2.7) are removed from the computation since they are no 
longer bound to the cluster and are hence considered lost from the 
cluster instantly on the relaxation timescale.
The method allows stars to have arbitrary masses and makes it very 
easy to allow for a stellar mass spectrum in the calculations.

We now describe our implementation of the Monte-Carlo method in detail. 
For completeness, we also include some of the basic equations of
the method. For derivations of these equations, and a more detailed 
discussion of the basic method, see H\'enon (1971b), 
Stod\'olkiewicz (1982), and Spitzer (1987).

\subsection{Initial Conditions}

The initial model is assumed to be in dynamical equilibrium, 
so that the potential does not change on the crossing timescale. 
This is important since the Monte-Carlo method uses a 
timestep which is of the order of the relaxation time, and hence cannot 
handle the initial phase of ``violent relaxation'' during which the potential 
changes on the dynamical timescale. Under the assumption of spherical symmetry,
the distribution function for such an equilibrium system can be written in the 
form $f = \Psi(E, J)$, where $E$ and $J$ are the energy per unit mass, 
and angular momentum per unit mass,

\begin{eqnarray}
E & = & \Phi(r) + \frac{1}{2}(v_r^2 + v_t^2) \rm{,} \\
J & = & r v_t.
\end{eqnarray} 

\noindent
Here $r$ is the distance from the cluster center, $v_r$ is the radial velocity, 
$v_t$ is the transverse velocity,  and $\Phi(r)$ is the gravitational potential. 
In principle, the initial distribution 
function $\Psi(E,J)$ can be arbitrary. However, in practice, computing
a self-consistent potential for an arbitrary distribution function can be quite
difficult. Since the method requires the initial potential $\Phi(r)$ to be known, 
a simple initial model is usually selected so as to allow the potential to be 
computed quasi-analytically. Common examples are the sequence of King models and
the Plummer model.

Once the number of stars $N$ is selected, the initial condition is constructed 
by assigning to each star values for $r$, $v_r$, $v_t$, and $m$, consistent 
with the selected model. Once the positions
and masses of all the stars are known, the gravitational potential $\Phi$ is computed
as a function of distance from the center. The energy per unit mass $E$,
and angular momentum per unit mass $J$ of each star are then computed using 
equations~(1) and (2).

\subsection{The Gravitational Potential}

We compute the mean potential of the cluster by summing the potential due to each star,
under the assumption of spherical symmetry.
We use only the radial position $r$ of each star (since we assume spherical symmetry,
we can neglect the angular positions of the stars, to a very good approximation).
We begin by sorting all the stars by increasing radius. Then the 
potential at a point $r$, which lies between two stars at positions $r_k$ and $r_{k+1}$,
is given by

\begin{eqnarray}
\Phi(r) & = & G\left(-\frac{1}{r}\sum_{i=1}^{k}{m_i} \, - \sum_{i=k+1}^{N}{\frac{m_i}{r_i}} \right).
\end{eqnarray} 

For any two neighboring stars at distances $r_k$ and $r_{k+1}$, the mass contained within the 
radius $r$ remains constant for $r_k < r < r_{k+1}$. Hence, we can compute the potential at r, 
if the potentials 
$\Phi_k = \Phi (r_k)$ and $\Phi_{k+1} = \Phi (r_{k+1})$ are known, as
\begin{eqnarray}
\Phi(r) & = & \Phi_k + \left(\frac{1/r_k - 1/r}{1/r_k - 1/r_{k+1}}\right)
	\left(\Phi_{k+1} - \Phi_k\right).
\end{eqnarray}

At each timestep, we store pre-computed values of $\Phi_k = \Phi(r_k)$, for each star k in
the cluster. The potential at an arbitrary point $r$ can then be quickly computed simply by
finding the index k such that $r_k \le r \le r_{k+1}$ and then using equation~(4).

We now describe the process of evolving the system through one complete timestep. 

\subsection{Two-Body Relaxation and Timestep Selection}

We simulate the effect of interactions during each timestep $\Delta t$ by 
perturbing the energy and angular momentum of each star in the cluster. The perturbations
$\Delta E$ and $\Delta J$ for a star are determined by computing a single \emph{effective} 
encounter between the star and its nearest neighbor (in terms of distance from the center, 
since we assume spherical symmetry). During such an encounter, the two stars exchange kinetic
energy, but the total energy is conserved. In the center of mass frame of the two
interacting stars, the magnitude of the velocity does not change; instead the velocity
is deflected through an angle $\beta$. 

In the original method described by H\'enon (1971b), the timestep used was a 
small fraction of the relaxation time for \emph{the entire cluster}. Although
the timestep computed in this way is suitable for the outer regions of the 
cluster, it is too large to provide an accurate representation of the relaxation
in the core, especially in the later stages of cluster evolution where the
relaxation time in the core can be many orders of magnitude smaller than
in the outer regions. This caused the inner regions of the cluster to be
under-relaxed. The limited computational resources available at that time 
did not permit the timestep to be made much smaller, without slowing down
the computation to a crawl. The greatly increased computational power available 
today allows us to use a timestep that is small enough to resolve the relaxation
process in the core, even for systems with $N \go 10^5$.

To provide an accurate description of the overall relaxation of the cluster, 
each effective encounter should give the correct mean value of the change in energy at 
each position. 
We achieve this by selecting the effective deflection angle $\beta_e$ for the encounter 
(in the center of mass frame of the two interacting stars) as follows.
If the masses of the two stars are $m_1$ and $m_2$, and their velocities $v_1$ and $v_2$, 
respectively, then the kinetic energy changes can be written as 
\begin{eqnarray}
\Delta KE_1 &=& m_1 v_1 \Delta v_1 + \frac{1}{2} m_1 (\Delta v_1)^2,  \\
\Delta KE_2 &=& m_2 v_2 \Delta v_2 + \frac{1}{2} m_2 (\Delta v_2)^2,
\end{eqnarray}
\noindent where $\Delta v_1$ and $\Delta v_2$ are the changes in the velocities during the 
encounter. 
Since the total kinetic energy in each encounter is conserved, the mean
value of the first terms on the RHS of equations~(5) and (6) must equal 
the mean value of the second terms (with the opposite sign). This indicates
that in order to get a good representation of the energy exchange between 
stars in the relaxation process, we must consider the
mean value of $m_1 (\Delta v_1)^2$ during each timestep. 

The change in velocity $\Delta v_1$ during an
encounter with a deflection angle $\beta$, can be calculated from elementary 
mechanics as (see, e.g., Spitzer 1987, eq. [2-6]),
\begin{eqnarray}
(\Delta v_1)^2 = 4 \frac{m_2^2}{(m_1 + m_2)^2} w^2 \sin^2 (\beta / 2),
\end{eqnarray}
\noindent where $w$ is the relative speed of the two stars before the encounter.
The mean overall \emph{rate} of change in the velocity $<(\Delta v_1)^2>$ due to many 
distant (weak) encounters of the star with other cluster stars
can then be calculated by averaging over the impact parameter (cf. Spitzer 1987, eq. [2-8]).
Using this, the mean change in the velocity in the time $\Delta t$ is given by
\begin{eqnarray}
<(\Delta v_1)^2> = 8\pi G^2 \nu \Delta t <m_2^2 w^{-1}> \ln \Lambda,
\end{eqnarray}
\noindent 
where $\ln \Lambda \equiv \ln (\gamma N)$ is the Coulomb logarithm
($\gamma$ is a constant $\sim 0.1$; see \S 3.1), and 
$\nu$ is the local number density of stars.
We obtain the correct mean value of $m_1 (\Delta v_1)^2$ by equating the RHS
of equations~(7) and (8), giving
\begin{eqnarray}
<4 \frac{m_1 m_2^2}{(m_1 + m_2)^2} w^2 \sin^2 (\beta / 2)> = 
8\pi G^2 \nu \Delta t <m_1 m_2^2 w^{-1}> \ln(\gamma N).
\end{eqnarray}
\noindent
Equation~(9) relates the timestep $\Delta t$ to the deflection angle $\beta$ 
for the encounter. Thus, in order to get the correct mean value of 
$m_1 (\Delta v_1)^2$ for the star during the time $\Delta t$, 
we can define the \emph{effective} deflection angle $\beta_e$ for the representative
encounter, as
\begin{eqnarray}
\sin^2(\beta_e/2) = 2\pi G^2 \frac{(m_1 + m_2)^2}{w^3} \nu \Delta t \ln(\gamma N).
\end{eqnarray}

In addition to using the correct mean value of $m_1 (\Delta v_1)^2$, we can also 
require that its variance be correct. To compute the variance, we must calculate
the mean value of $(\Delta v_1)^4$. Using equation~(7), we have
\begin{eqnarray}
(\Delta v_1)^4 = 16 \frac{m_2^4}{(m_1 + m_2)^4} w^4 \sin^4 (\beta/2).
\end{eqnarray}
\noindent
We then use Spitzer's equation~(2-5), and again integrate over the impact parameter
to get the mean value of $(\Delta v_1)^4$ in the time $\Delta t$, 
\begin{eqnarray}
<(\Delta v_1)^4> = 16 \pi G^2 \frac{m_2^4}{(m_1 + m_2)^2} w \nu \Delta t.
\end{eqnarray}
Comparing equations~(11) and (12), we see that, in order to have the correct 
variance of $m_1 (\Delta v_1)^2$, we should have
\begin{eqnarray}
\sin^4(\beta_e/2) = \pi G^2 \frac{(m_1 + m_2)^2}{w^3} \nu \Delta t.
\end{eqnarray}
Consistency between equations~(10) and (13) gives the relation between the number 
of stars in the system, and
the effective deflection angle that must be used,
\begin{eqnarray}
\sin^2(\beta_e/2) = \frac{1}{2 \ln (\gamma N)}.
\end{eqnarray}
\noindent
This relation indicates that for large $N$, the effective deflection angle must be 
small, while as $N$ decreases, close encounters become more important.
If the timestep is too large, then $<\sin^2(\beta /2)>$ is also too large, and the
system is under-relaxed. Hence the timestep used should be sufficiently small 
so as to get a good representation of the relaxation process in the cluster.
In addition, the local relaxation time varies greatly with distance from 
the cluster center. In practice we use the shortest relaxation time in the core to 
compute the timestep. We first evaluate the local density $\rho_c$ in the core and
the approximate core radius $r_c = (3 v_c^2/4 \pi G \rho_c)^{1/2}$. We then compute
the timestep $\Delta t$ using equation~(10) and requiring that the average value of 
$\sin^2(\beta_e/2)$ for the stars within the core radius $r_c$ be 
sufficiently small. 
The value of $\sin^2(\beta_e/2)$ given by equation~(14) 
varies only slightly between $0.046$ and $0.072$ for $N$ between 
$10^4$ and $5\times10^5$ (assuming $\gamma \simeq 0.1$). Hence for 
all our simulations, we require that $\sin^2(\beta_e/2) \lo 0.05$.

Equation~(10) is then used to compute the effective deflection angle 
for all stars in the cluster. The local number density $\nu$ is computed 
by averaging over the nearest $p$ stars.
We find that using a value of $p$ between 20 and 50 gives the best results 
for $N \sim 10^5$. We find that the difference in the core-collapse times 
obtained for various test models using values of $p$ between 20 and 50 
is less than 1\%. Of course,
the value of $p$ should not be too large so as to maintain a truly local 
estimate of the number density.  We use a value of $p=40$ in all our 
calculations, which gives consistently good agreement with published results.

\subsection{Computing the Perturbations $\Delta E$ and $\Delta J$ during an Encounter}

To compute the velocity perturbation during each timestep, a single representative 
encounter is computed for each star, with its nearest neighbor in radius. 
Selecting the nearest neighbor ensures that the correct local velocity distribution
is sampled, and also accounts for any anisotropy in the orbits. Due to spherical
symmetry, selecting the nearest neighbor in radius is equivalent to selecting
the nearest neighbor in 3-D, since only the velocity (and not the position) 
of the nearest neighbor is used in the encounter. 
Following H\'enon's notation, we
let ($r$, $v_r$, $v_t$) and ($r^{\prime}$, $v_r^{\prime}$, $v_t^{\prime}$)
represent the phase space coordinates of the two interacting stars, with masses $m$ and
$m^{\prime}$, respectively. In addition to these
parameters, the angle $\psi$ of the plane of relative motion defined by 
($\mathbf{r^{\prime}} - \mathbf{r}$, $\mathbf{v^{\prime}} - \mathbf{v}$)
with some reference plane is selected randomly between 0 and $2\pi$, since the 
distribution of field stars is assumed to be spherically symmetric.  

We take our frame of reference such that the $z$ axis is parallel to $\mathbf{r}$,
and the $(x,z)$ plane contains $\mathbf{v}$. Then the velocities of the two 
stars are given by
\begin{eqnarray}
{\mathbf v} = (v_t, 0, v_r), \; {\mathbf v^{\prime}} = 
(v_t^{\prime} \cos\phi, v_t^{\prime} \sin\phi, v_r^{\prime}),
\end{eqnarray}
\noindent where $\phi$ is also randomly selected between 0 and $2\pi$, since the 
transverse velocities are isotropic because of spherical symmetry. 
The relative velocity ${\mathbf w} = (w_x, w_y, w_z)$ is then
\begin{eqnarray}
{\mathbf w} = (v_t^{\prime}\cos \phi - v_t, v_t^{\prime}\sin \phi, v_r^{\prime} - v_r).
\end{eqnarray}

We now define two vectors ${\mathbf w_1}$ and ${\mathbf w_2}$ with the same magnitude
as ${\mathbf w}$, such that ${\mathbf w_1}$, ${\mathbf w_2}$, and ${\mathbf w}$ are
mutually orthogonal. The vectors ${\mathbf w_1}$ and ${\mathbf w_2}$ are given by
\begin{eqnarray}
{\mathbf w_1} &=& (w_y w / w_p, - w_x w / w_p, 0), \\
{\mathbf w_2} &=& (-w_x w_z/w_p, -w_y w_z/w_p, w_p),
\end{eqnarray}
\noindent where $w_p = (w_x^2 + w_y^2)^{1/2}$.
The angle $\psi$ is measured from the plane containing the vectors $\mathbf{w}$
and $\mathbf{w_1}$. The relative velocity of the two stars after the encounter
is given by
\begin{eqnarray}
{\mathbf w^{\star}} = {\mathbf w} \cos\beta + {\mathbf w_1} \sin\beta \cos\psi 
			+ {\mathbf w_2} \sin\beta \sin\psi,
\end{eqnarray}
where $\beta$ is the deflection angle computed in \S 2.4.
The new velocities of the two stars after the interaction are then given by
\begin{eqnarray}
{\mathbf v^{\star}} &=& {\mathbf v} - \frac{m^{\prime}}{m + m^{\prime}} 
			({\mathbf w^{\star}} - {\mathbf w}),  \\
\mathbf{v^{\prime\star}} &=& {\mathbf v^{\prime}} + \frac{m}{m + m^{\prime}} 
			({\mathbf w^{\star}} - {\mathbf w}). 
\end{eqnarray}

The new radial and transverse velocities for the first star are given by 
$v_r^{\star} = v_z^{\star}$, and 
$v_t^{\star} = (v_x^{\star 2}+v_y^{\star 2})^{1/2}$, from which we compute
the new orbital energy $E$ and angular momentum $J$ as
$E^{\star} = \Phi(r) + \frac{1}{2}(v_r^{\star 2}+v_t^{\star 2})$, and 
$J^{\star} = r v_t^{\star}$. Similar quantities $E^{\prime\star}$ and 
$J^{\prime\star}$ are also computed for the second star.

\subsection{Computing New Positions and Velocities}

Once the orbits of all the stars are perturbed, i.e., new values of 
$E$ and $J$ are computed for each star, a new realization of the system is
generated, by selecting a new position for each star in its new orbit, in such a way
that each position in the orbit is weighted by the amount of time that the star spends
at that position. To do this, we begin by computing the pericenter and apocenter 
distances, $r_{min}$ and $r_{max}$, for each star. The orbit of a star 
in the cluster potential is a rosette, with $r$ oscillating between 
$r_{min}$ and $r_{max}$, which are roots of the equation
\begin{eqnarray}
Q(r) = 2 E - 2 \Phi(r) - J^2/r^2 = 0.
\end{eqnarray}
See Binney \& Tremaine (1987; \S 3.1) for a general discussion, and see
H\'enon (1971b; Eqs.~[41]-[45]) for a convenient method of solution.
The new position $r$ should now be selected between $r_{min}$ and $r_{max}$, in such a way 
that the probability of finding r in an interval $dr$ is equal to the fraction of
time spent by the star in the interval during one orbit, i.e.,
\begin{eqnarray}
\frac{dt}{P} = \frac{dr/|v_r|}{\int_{r_{min}}^{r_{max}}{dr/|v_r|}}, 
\end{eqnarray}
\noindent where $P$ is the orbital period, and $|v_r|$ is given by
\begin{eqnarray}
|v_r| = [2 E - 2 \Phi(r) - J^2/r^2]^{1/2} = [Q(r)]^{1/2}.
\end{eqnarray}

Thus the value of r should be selected from a probability distribution that is
proportional to $f(r) = 1/|v_r|$. Unfortunately, at the pericenter and apocenter points
($r_{min}$ and $r_{max}$), the radial velocity $v_r$ is zero, and the probability distribution 
becomes infinite. To overcome this problem, we make a change of coordinates
by defining a suitable function $r = r(s)$ and selecting a value of s from the distribution
\begin{eqnarray}
g(s) \equiv \frac{1}{|v_r|} \frac{dr}{ds}.
\end{eqnarray}
We must select the function $r(s)$ such that $g(s)$ remains finite in the 
entire interval. A convenient function $r(s)$ that satisfies these requirements 
is given by 
\begin{eqnarray}
r = \frac{1}{2}(r_{min} + r_{max}) + \frac{1}{4}(r_{max} - r_{min})(3s - s^3), 
\end{eqnarray}
\noindent where $s$ lies in the interval -1 to 1. 
We then generate a value for $s$, which is consistent with the distribution $g(s)$,
using the von Neumann rejection technique. Equation~(26) then gives a corresponding value
for $r$ which is consistent with the distribution function $f(r)$.

The magnitude of the new radial velocity $v_r$ is computed using equation~(24), and 
its sign is selected randomly. The transverse velocity is given by $v_t = J/r$.

Once a new position is selected for each star using the above procedure, the gravitational
potential $\Phi(r)$ is recomputed as described in \S 2.3. This completes the timestep, 
and allows the next timestep to be started.

Note that the gravitational potential used to compute new positions and velocities
of the stars is from the previous timestep. The new potential can only be computed 
\emph{after} the new positions are assigned, and it is then used to recompute the 
positions in the next timestep. Thus the computed potential always lags slightly 
behind the actual potential of the system. The exact potential is known only at 
the initial condition. This only introduces a small systematic error in the 
computation, since the potential changes significantly only on the relaxation 
timescale. 

A more important source of error, especially in computing the new energies of the stars
after the potential is recomputed, is the random fluctuation of the potential in the
core, which contains relatively few stars, but has a high number density.
Since the derivative of the potential is also steepest in the core, a small error in 
computing a star's position in the core can lead to a large error in computing its
energy. As the simulation progresses, this causes a slow but consistent leak in the 
total system energy. The magnitude of this error (i.e., the amount of energy lost
per timestep) depends partly on the number of stars $N$ in the system.
For large $N$, the grid on which the potential is pre-computed (see \S 2.3) is finer,
and the number of stars in the core is larger, which reduces the noise in the 
potential. The overall error in energy during the course of an entire simulation 
is typically of order a few percent for $N = 10^5$ stars. 
In any realistic simulation, the actual energy 
gain or loss due to real physical processes such as stellar evolution, escape of stars 
through a tidal boundary, and interactions involving binaries, is at least an order of 
magnitude greater than this error. Hence we choose not to renormalize the energy of 
the system, or employ any other method to artificially conserve the energy 
of the system, which could affect other aspects of the evolution.

Another possible source of error in Monte-Carlo simulations, 
which was noted by H\'enon (1971b) is the ``spurious relaxation'' effect.
This is the tendency for the system to relax because of the potential 
fluctuations from one timestep to the next, even in the absence of orbital
perturbations due to two-body relaxation. However, this effect
is significant only for simulations done with very low $N \sim 10^2-10^3$.
In test calculations performed with $N \sim 10^4-10^5$ and two-body relaxation 
explicitly turned off (by setting the scattering angle $\beta_e = 0$ in eq.~[10]), 
we find no evidence of spurious relaxation. 
Indeed H\'enon (1971b) himself showed that spurious relaxation was not
significant in his models for $N \go 10^3$. 

\subsection{Escape of Stars and the Effect of a Tidal Boundary}

For an isolated system, the gradual evaporation of stars from the cluster is computed
in the following way. 
During each timestep, after the perturbations  $\Delta E$ and $\Delta J$ are 
computed, all stars with a positive total energy (given by eq.~[1]) are assumed to 
leave the cluster on the crossing timescale. They are therefore considered lost 
immediately on the relaxation timescale, and removed from the simulation. The
mass of the cluster (and its total energy) decreases gradually as a result of this
evaporation process.

As a simple first step to take in to account the tidal field of the Galaxy, 
we include an effective tidal boundary around the cluster,
at a distance $r_t \simeq R_g (M_{cluster}/3 M_g)^{1/3}$, where $R_g$ is the distance
of the cluster from the Galactic center and $M_g$ is the mass of the Galaxy
(approximated as a point mass). The tidal radius is roughly the size of the
Roche lobe of the cluster in the field of the Galaxy. Once the initial tidal 
radius ${r_t}_0$ is specified, the tidal radius at a subsequent time $t$ during 
the simulation can be computed by 
$r_t(t) = {r_t}_0 (M_{cluster}(t)/M_{cluster}(0))^{1/3}$. After each timestep,
we remove all stars with an apocenter distance $r_{max}$ greater than the 
tidal radius, since they are lost from the cluster on 
the crossing timescale. As the cluster loses stars due to evaporation and
the presence of the tidal boundary, its mass decreases, which causes the tidal
boundary to shrink, in turn causing even more stars to be lost. 
The total mass loss due to a tidal boundary can be very significant,
causing up to 90\% of the mass to be lost (depending on the initial model) 
over the course of the simulation (see \S 3.2).

\subsection{Units}

Following the convention of most previous studies, we define dynamical units
so that $[G] = [M_0] = [-4 E_0] = 1$, where $M_0$ and $E_0$ are
the initial total mass and total energy of the system (H\'enon 1971).
Then the units of length $L$, and time $T$ are given by
\begin{eqnarray}
L = G M_0^2 (-4 E_0)^{-1} {\rm ,} \hspace{.5in} {\rm and}
\hspace{.25in} T = G M_0^{5/2} (-4 E_0)^{-3/2} {\rm .}  
\end{eqnarray}
\noindent
We see that $L$ is basically the virial radius of the cluster, and $T$ 
is of the order of the initial dynamical (crossing) time. 
To compute the evolution of the cluster on a relaxation timescale, 
we rescale the unit of time to $T N_0/\ln (\gamma N_0)$, which is of the order 
of the initial relaxation time. Using this unit of time
allows us to eliminate the $\ln (\gamma N)$ dependence of the evolution
equations. The only equation that explicitly contains
the evolution time is equation~(10), which relates the timestep and the effective
deflection angle. In our units, equation~(10) can be written as,
\begin{eqnarray}
\left[\sin^2(\beta_e/2)\right] = 2\pi \frac{([m_1] + [m_2])^2}{[w]^3} [\nu] [\Delta t] N,
\end{eqnarray}
\noindent where [q] indicates a quantity q expressed in our simulation units.
Using a unit of time that is proportional to the initial relaxation time
has the advantage that the evolution timescale is roughly independent of
the number of stars $N$ once an initial model has been selected.
This is only true approximately, for isolated systems 
of equal-mass stars, with no other processes that depend explicitly on the 
number of stars (such as stellar evolution or mass segregation). 
For example, the half-mass relaxation time for the 
Plummer model, 
\begin{eqnarray}
t_{rh} = \frac{0.138 N}{\ln (\gamma N)} \left(\frac{r_h^3}{GM} \right)^{1/2},
\end{eqnarray}
\noindent is always 0.093 in our units, independent of $N$. 

The dynamical units defined above are identical to the standard $N$-body units 
(Heggie \& Mathieu 1986). Hence to convert the evolution 
time from $N$-body time units to our Monte-Carlo units, we must simply
multiply by a factor $\ln (\gamma N_0) / N_0$.

\subsection{Numerical Implementation}

We have implemented our Monte-Carlo code on the SGI/CRAY Origin2000 parallel
supercomputer at the National Center for Supercomputing Applications (NCSA),
and at Boston University.
Our parallelized code can be used to get significant speedup of the 
simulations, using up to 8 processors, especially for large $N$ simulations. 
This ability to perform large $N$ simulations
will be particularly useful for doing realistic simulations of very large
globular clusters such as 47 Tuc (with $N \go 10^6$ stars).
A simulation with $N = 10^5$ stars can be completed in approximately 
15--20 CPU hours on the Origin2000, 
which uses MIPS R10000 processors. For comparison, a simulation of this 
size would take $\sim 6$ months to complete using the GRAPE-4, which is 
the fastest available hardware for $N$-body methods.

The most computationally intensive step in the simulation is the calculation
of the new positions of stars. The operation involves solving for the roots of 
an equation (eq.~[22]) using the indexed values of the positions 
of the $N$ stars. 
We find that the most efficient method to solve for the roots in this case 
is the simple bisection method (e.g., Press \etal 1992), which requires 
$\sim N \log_2 N$ steps to converge to the root. 
Hence the computation of the positions and velocities also scales as 
$\sim N \log_2 N$ in our method. The next most expensive operation is the
evaluation of the potential at a given point $r$. As described in \S 2.3, 
this requires finding $k$ such that $r_k \le r \le r_{k+1}$ and then 
using equation~(4). This search can again be done easily using the bisection
algorithm. However, since the evaluation of the potential is required
several times for each star, in each timestep, it is useful to tabulate
the values of $k$ on fine grid in $r$ at the beginning of the timestep.
This allows the required values of $k$ to be found very quickly, at
the minor cost of using more memory to store the table. The rest of the
steps in the simulation scale almost linearly with $N$. This makes the
overall computation time scale (theoretically) as $N \log_2 N$.

In Figure~1, we show the scaling of the wall-clock time with the number of 
processors, and also the scaling of the overall computation time with the
number of stars $N$ in the simulation. The overall computation time is 
consistent with the theoretical estimate for $N \lo 10^5$. For
larger $N$, the computation time is significantly higher, because of the
less efficient use of cache memory and other hardware inefficiencies
that are introduced while handling large arrays. For $N$ in the
range $1-5\times 10^5$,
we find that the actual computation time scales as $\sim N^{1.4}$.

We find that we can easily reduce the overall computation time by a factor
of $\approx 3$ by using up to 8 processors. The scaling is most efficient for $2-4$ 
processors for simulations with $N \sim 1-5\times 10^5$. The scaling
gets progressively worse for more than 8 processors. This is in part caused by
the distributed shared-memory architecture of the Origin2000 supercomputer, which
allows very fast communication between the nearest 2-4 processors, but slower 
communication between the nearest 8 processors. Beyond 8 processors, the 
communication is even slower, since the processors are located on different 
nodes. The most suitable architecture for implementing the parallel Monte-Carlo
code would be a truly shared memory supercomputer, with roughly uniform memory 
access times between processors.
Our code is implemented using the Message Passing Interface (MPI) parallelization
library, which is actively being developed and improved. The MPI standard
is highly portable, and available on practically all parallel computing platforms
in use today. The MPI library is optimized for each platform and automatically
takes advantage of the memory architecture to the maximum extent possible.
Hence we expect that future improvements in the communication speed and memory 
architectures will make our code scale even better. We are also in the 
process of improving the scaling of the code to a larger number of processors
by designing a new algorithm for reducing the amount of communication required
between processors. This will be described in detail in a subsequent paper,
where we incorporate primordial binary interactions in our code.

\section{Test Results}

In this section, we describe our first results using the new Monte-Carlo code 
to compute the evolution of the Plummer and King models. We explore the
evolution of the Plummer model in detail, and compare our results with 
those obtained using Fokker-Planck and $N$-body methods. We also compare 
core-collapse times and mass-loss rates for the series of King 
models ($W_0 = 1-12$), including a tidal radius, with similar results 
obtained by Quinlan (1996) using a 1-D Fokker-Planck method.

\subsection{Evolution of an Isolated Plummer Model}

We first consider the evolution of a cluster with the Plummer model (which is a 
polytropic model, with index $n=5$; see, e.g., Binney \& Tremaine 1987) as the
initial condition. 
Perhaps the best known result for single component systems, 
is the expected homologous evolution of the halo, leading to the eventual 
development of a power-law density profile between the core and the outer
halo, during the late phases of evolution.
At late times the cluster evolves through a sequence of nearly self-similar
configurations, with the core contracting and a power-law halo with
density $\rho \propto r^{-\beta}$ expanding out.
The development of this power law has been predicted theoretically 
(Lynden-Bell \& Eggleton 1980; Heggie \& Stevenson 1988), and verified 
using direct Fokker-Planck integrations (Cohn 1980). 
The exponent $\beta$ is theoretically and numerically estimated 
to be about $2.2$ (Spitzer 1987). 
However, since the theoretical derivations are based on an analysis of the 
Fokker-Planck equation, it is not surprising that the numerical Fokker-Planck 
integrations (which solve the same Fokker-Planck equation numerically) reproduce 
the theoretical exponent exactly. Due to limitations in computing accurate density 
profiles using a small number of stars, this result has not been confirmed 
independently using an $N$-body simulation. 

Here, we explore numerically for the first time the development of this power law
using an independent method. Some early results were obtained using 
previous versions of the Monte-Carlo method, but with a small number of stars 
$N \sim 10^3$ (Duncan \& Shapiro 1982). Although the Monte-Carlo
method can be thought of as just another way of solving the Fokker-Planck equation, 
there are significant differences between solving the equation in the 
continuous limit ($N \to \infty$), as in direct Fokker-Planck integrations, 
and by using a discrete system with a finite $N$ as in our method. There
are also many subtle differences in the assumptions and approximations 
made in the two methods, and even in different implementations of the same
method.

In Figures~2a--c we show the density profile of the cluster at three different
times during its evolution, up to core collapse. We start with an $N = 10^5$
isolated Plummer model, and follow the evolution up to core-collapse, which
occurs at $t = t_{cc} \simeq 15.2\, t_{rh}$. This simulation, performed with 
$N = 10^5$ stars, took about 18 CPU hours on the SGI/Cray Origin2000.
In our calculations, the core-collapse time is taken as the time when the
the innermost lagrange radius (radius containing 0.3\% of the
total mass of the cluster) becomes smaller than 0.001 (in our units described
in \S 2.8), at which point the simulation is terminated. Given the very rapid
evolution of the core near core collapse, we find that we can determine the
core-collapse time to within $\lo 1\%$. The accuracy is limited mainly by
noise in the core. 
The value we obtain for  $t_{cc} / t_{rh}$ is in very good 
agreement with other core-collapse times between $15-16 \, t_{rh}$ for 
the Plummer model, reported using other methods. 
For example Quinlan (1996) obtains a core collapse time of 
$15.4 \, t_{rh}$ for the Plummer model using a 1-D Fokker-Planck method, and
Takahashi (1993) finds a value of $15.6 \, t_{rh}$, using a variational method 
to solve the 1-D Fokker-Planck equation. 

Figure~2a shows the density profile at an intermediate time $t = 11.4\, t_{rh}$
during the evolution. The dotted line indicates the initial Plummer profile.
At this point in the evolution, we still see a well defined core, with
the core density increased by a factor of $\sim 30$ compared to the initial 
core density. We see the power-law density profile developing, with the
best-fit index $\beta = 2.8$. 
In Figure~2b, we show the density profile just before core collapse,
at $t = 15\, t_{rh}$. We see that the core density has now increased
by a factor of $\sim 10^4$ over the initial core density. The power
law is now clearly visible, with the best-fit index $\beta = 2.3$.
Finally, in Figure~2c, we show the density profile at core-collapse,
$t = 15.2\, t_{rh}$. The dashed line now indicates the \emph{theoretical}
power law with $\beta = 2.2$. We see that the actual density profile
seems to approach the theoretical profile asymptotically as the 
system approaches core collapse. 
At this point in the evolution, the core density as measured in our 
simulation is about $10^6$ times greater than the initial density.
In a globular cluster with $N = 2 \times 10^5$, an average 
stellar mass $<m> = 0.5 \, \rm{M_{\odot}}$, and a mean velocity 
dispersion $<v^2>^{1/2} = 5 \, \rm{km\,s^{-1}}$, this would correspond 
to a number density of $\sim 2 \times 10^9\, \rm{pc^{-3}}$.
Note that a real globular cluster is not expected to reach such high 
core densities, since the formation of binaries and the subsequent 
heating of the core due to binary interactions become significant 
at much lower densities. 
Numerical noise due to the extremely small size of the core makes it
difficult to determine the core radius and density accurately at this stage. 
This also causes the numerical accuracy of the Monte-Carlo method to
deteriorate, forcing us to stop the computation. 
Thus, we find that the power-law structure of the density profile
as the cluster approaches core collapse is consistent with  
theoretical predictions, and the power-law index approaches
its theoretical value asymptotically during the late stages of 
core collapse. 

Next, we look at the evolution of the Lagrange radii (radii containing 
constant fractions of the total mass), and we compare our results
with those of an equivalent $N$-body simulation.  
In Figure~3, we show the evolution of the Lagrange radii 
for an $N=16384$ direct $N$-body integration by Makino (1996) 
and for our Monte-Carlo integration with $N=10^5$ stars.
Time in the direct $N$-body integration is scaled to the initial 
relaxation time (the standard time unit in our Monte Carlo method)
using equation~(27) with $\gamma = 0.11$ (see Heggie \& Mathieu 1986; 
Giersz \& Heggie 1994; Makino 1996).
The agreement between the $N$-body and Monte Carlo results
is excellent over the entire range of Lagrange radii and time.  
The small discrepancy in the outer Lagrange radii is caused in part
by a different treatment of escaping stars in the two models. 
In the Monte Carlo model, escaping stars are removed from the 
simulation and therefore not included in the determination of the 
Lagrange radii, whereas in the $N$-body model escaping stars are 
not removed. The difference is further explained by the effect of strong 
encounters, which is greater in the $N$-body simulation 
by a factor $\sim \ln(10^5)/\ln(16384)$, or about 20\%.
In an isolated cluster, the overall evaporation rate
is very low (less than 1\% of stars escape up to core collapse).
In this regime, the escape of stars is dominated by strong 
interactions in the core. Since the orbit-averaged Fokker-Planck 
equation is only valid when the fractional energy change per orbit 
is small, it does not account for strong interactions.
Hence, our Monte-Carlo simulations cannot accurately predict 
the rate of evaporation from an isolated cluster 
(see, e.g., Binney \& Tremaine 1987, \S 8.4). 
This problem does not occur in tidally truncated clusters, where
the escape rate is much higher, and is dominated by the diffusion 
of stars across the tidal boundary, and not by strong interactions.

In Figure~4 we show the evolution of various global quantities for 
the system during the same simulation as in Figure~3. The virial ratio 
($K/|W|$, where $K$ and $W$ are the total kinetic and potential
energies of the cluster) remains very close to 0.5 (within 1\%), 
indicating that dynamical equilibrium is 
maintained very well during the entire simulation. The virial ratio provides 
a very good measure of the quality of our numerical results, since it 
is not controlled in our calculations (except for the initial model, which 
is constructed to be in equilibrium). We see that in the absence of a tidal radius, 
there is very little mass loss (less than 1\%), and hence very little energy 
is carried away by escaping stars.

\subsection{Evolution of Isolated and Tidally Truncated King models}

King models (King 1966) have long been used to fit observed profiles of 
globular clusters. They usually provide a very good fit for most clusters, 
except for those which have reached core collapse. A King model 
has a well-defined, nearly constant-density core, and a ``lowered Maxwellian''
velocity distribution, which represents the presence of a finite tidal radius.
A King model is usually specified in terms of the dimensionless 
central potential $W_0$ or, equivalently, the central concentration 
$c = \log(r_t/r_c)$, where $r_t$ is the tidal radius, and $r_c$ is the core radius. 

We study the evolution of the entire family of King models from 
$W_0 = 1$ to $W_0 = 12$, in two different configurations. 
We first consider the evolution of an isolated cluster 
i.e., even though the initial King model is truncated at its finite tidal 
radius, we do not enforce that tidal boundary during the evolution, allowing the 
cluster to expand indefinitely. 
We compute the core-collapse times for the entire sequence of King models. 
We then redo the calculations with a tidal boundary
in place, to determine the enhanced rate of mass loss from the cluster and
the final remaining mass at the time of core collapse. 
We compare our results for the sequence of King models with equivalent results
obtained by Quinlan (1996) using direct Fokker-Planck integrations in 1-D.
In Table~1, we show the core collapse times for the various models, along with
the equivalent results from Quinlan (1996). All our Monte-Carlo calculations were
performed using $N = 10^5$ stars. We see that the agreement in the core collapse times 
for isolated clusters is excellent (within a few percent for the low-$W_0$ models, 
and within 10\% up to $W_0 = 9$). 
For $W_0 > 9$, the agreement is still good, considering that the models start off in a 
highly collapsed state and therefore have very short core-collapse times, which 
leads to larger fractional errors. 

In Figure~5, we show the evolution of the Lagrange radii for a tidally
truncated King model with $W_0 = 3$. 
The initial tidal radius is $\simeq 3.1$ times the virial radius.
In this case, the mass loss through the tidal boundary is very
significant, as is seen from the evolution of the outer Lagrange radii. 
The mass loss causes the tidal radius to constantly move inward, which 
further accelerates the process. 
Figure~6 shows the evolution of the total mass and energy of the tidally
truncated cluster. Only 44\% of the initial mass is retained in the cluster
at core-collapse. Also, the binding energy of the cluster is significantly
lower at core-collapse, since the escaping stars carry away mass as well
as kinetic energy from the cluster.
In contrast, the evolution of an 
isolated $W_0 = 3$ King model is very much like that of the isolated
Plummer model described earlier, with a very low mass loss rate, and a 
longer core-collapse time of $t_{cc} = 17.7\,t_{rh}$ (in excellent
agreement with the value of $17.6\,t_{rh}$ computed by Quinlan 1996).

Our results for clusters with a tidal boundary show systematic differences 
from the 1-D Fokker-Planck results of Quinlan (1996).
We find that the mass loss through the tidal boundary is significantly 
higher for the low-concentration models ($W_0 < 6$) in the Fokker-Planck 
models. For the high-concentration ( $W_0 > 6$) models, the difference 
between isolated models and tidally truncated models is small, and 
the agreement between the methods remains very good. Hence, for low $W_0$, 
our models undergo core collapse at a much later time compared to the 
Fokker-Planck models, and retain more mass at core collapse.
This discrepancy is caused by the 1-D nature of the Fokker-Planck 
models. In 1-D Fokker-Planck calculations, stars
are considered lost from the cluster when their energy is greater than the
energy at the tidal radius. This clearly provides an overestimate of the
escape rate, since it assumes the most extended radial orbits for stars,
and ignores stars on more circular orbits with high angular 
momentum, which would have much smaller orbits at the same energy. 
In contrast, in the Monte-Carlo
method, the orbit of each star is computed using its energy \emph{and} angular
momentum, which allows the apocenter distance to be determined correctly. 
Stars are considered lost only if their apocenter distances from the 
cluster center are greater than the tidal radius.
As stars on radial orbits are removed preferentially, this 
creates an anisotropy within the cluster, which affects the overall evolution.
The artificially high rate of mass loss in 1-D Fokker-Planck simulations
has also been pointed out recently in comparisons with $N$-body results
(Portegies Zwart \etal 1998; Takahashi \& Portegies Zwart 1999). These
authors show that, with appropriate modifications, the results of 2-D
Fokker-Planck calculations can be made to agree much better with those from 
$N$-body simulations.
Indeed, we find that our result for the $W_0 = 3$ model with a tidal boundary 
($t_{cc} = 12.0\,t_{rh}$, and $M_{final} = 0.44$) agrees much better with that 
obtained using the improved 2-D Fokker-Planck method, which gives 
$t_{cc} = 11.3\,t_{rh}$, and $M_{final} = 0.34$ 
(Takahashi 1999, private communication).
For further comparison, and to better understand the cause of the higher mass
loss in the 1-D Fokker-Planck calculation, we have performed a Monte-Carlo 
simulation using the same energy-based escape criterion that is used in the 
1-D Fokker-Planck integrations. We find that using the energy-based escape
criterion for $W_0 = 3$ gives $t_{cc} = 10.9\,t_{rh}$, and $M_{final} = 0.30$, 
which agrees better with the 1-D Fokker-Planck result, but a significant 
discrepancy still remains. This is not surprising, since, even when using a
1-D escape criterion, our underlying method still remains 2-D.
Again, our result agrees better with the corresponding result obtained by 
Takahashi (1999, private communication) using the energy-based escape 
criterion in his 2-D Fokker-Planck method, $t_{cc} = 10.2\,t_{rh}$, 
and $M_{final} = 0.28$.
It is reassuring to note that the differences between our 2-D results and 
1-D Fokker-Planck results are also mirrored in the 2-D Fokker-Planck 
calculations of Takahashi.
Since our Monte-Carlo method is intrinsically 2-D, it is not possible for us
to do a true 1-D (isotropic) calculation to compare results directly with 
1-D Fokker-Planck calculations. 


\section{Summary and Future Directions}

We have presented results obtained using our new Monte-Carlo code for 
the evolution of clusters containing $10^5$ stars, up to core collapse. 
We have compared our results with those of 1-D Fokker-Planck calculations 
(Quinlan 1996) for isolated as well as tidally truncated King models with 
$W_0 = 1-12$. We find very good agreement for the core-collapse times of
isolated King models. For tidally truncated models (especially for $W_0 < 6$), 
we find that the escape rate of stars in our models is significantly lower than 
in the 1-D Fokker-Planck models. This is to be expected, since the 1-D
Fokker-Planck models use an energy-based escape criterion, which does not 
account for the anisotropy in the orbits of stars,
and hence overestimate the escape rate. This effect is most evident in
tidally truncated clusters, since stars on radial orbits are preferentially
removed, while those on more circular orbits (with the same energy) are not.
In one case ($W_0 = 3$), we have verified
that our results are in good agreement with those from new 2-D Fokker-Planck
calculations (Takahashi 1999, private communication), which properly account
for the velocity anisotropy, and use the same apocenter-based escape criterion
as in our models. Further comparisons of our results with 2-D Fokker-Planck
calculations will be presented in a subsequent paper (Joshi, Nave, \& Rasio 1999).
Our detailed comparison of the evolution of the Plummer model with an equivalent 
direct $N$-body simulation also shows excellent agreement between the two 
methods up to core collapse. 

Our results clearly show that the Monte-Carlo method provides a robust,
scalable and flexible alternative for studying the evolution of globular clusters.
Its strengths are complementary to those of other methods,
especially $N$-body simulations, which are still prohibitively expensive 
for studying large systems with $N \go 10^5$. The Monte-Carlo method
requires more computational resources compared to Fokker-Planck methods, 
but it is several orders of magnitude faster than $N$-body simulations.
The star-by-star representation of the system in this method makes it 
particularly well suited for studying the evolution of interesting 
sub-populations of stars within globular clusters, such as  
pulsars, blue stragglers, or black holes.

Our method also presents the interesting possibility of performing hybrid 
simulations that use the Monte-Carlo method for the bulk of the evolution 
of a cluster up to the core collapse phase, and then switch to an $N$-body 
simulation to follow the complex core-collapse phase during which the high 
reliability of the $N$-body method is desirable. The discreteness of the 
Monte-Carlo method, and the fact that it follows the same phase space parameters 
for a cluster as the $N$-body method, make it easy to switch from one method to
the other during a single simulation. 

In subsequent papers, we will present results for the dynamical evolution
of clusters with different 
mass spectra, including the effects of mass loss due to stellar evolution. 
We are also in 
the process of incorporating primordial binaries in our Monte-Carlo code, 
in order to follow the evolution in the post-core collapse phase. 
Dynamical interactions involving binaries will be treated using a
combination of direct numerical integrations of the orbits on a case-by-case 
basis and precomputed cross-sections. 
The cross-sections will be obtained from separate sets of scattering
experiments as well as fitting formulae (Sigurdsson \& Phinney 1995;
Heggie, Hut, \& McMillan 1996, and references therein).
Our code will also incorporate a simple treatment
of stellar evolution in binaries, using an extensive set of approximate recipes
and fitting formulae developed recently for STARLAB (Portegies Zwart 1995).
Simulations of clusters containing realistic numbers of stars {\it and\/}
binaries will allow us for the first time ever to compute detailed predictions
for the properties and distributions of all interaction products, including
blue stragglers (from mergers of main-sequence stars), X-ray binaries and
recycled pulsars (from interactions involving neutron stars) and cataclysmic
variables (from interactions involving white dwarfs).

\acknowledgements

We thank Piet Hut and Stephen McMillan for many helpful
discussions.  We are grateful to Jun Makino and Koji Takahashi for
kindly providing valuable data and answering numerous questions.
This work was supported in part by NSF Grant AST-9618116 and NASA ATP 
Grant NAG5-8460. F.A.R.\ was supported in part by an Alfred P.\
Sloan Research Fellowship.  F.A.R. also thanks the Theory Division of
the Harvard-Smithsonian Center for Astrophysics for hospitality.  This
work was also supported by the National Computational Science Alliance
under Grant AST970022N and utilized the SGI/Cray Origin2000
supercomputer at Boston University.  NASA also supported this work
through Hubble Fellowship grant HF-01112.01-98A awarded (to SPZ) by
the Space Telescope Science Institute, which is operated by the
Association of Universities for Research in Astronomy, Inc., for NASA
under contract NAS\, 5-26555.  SPZ is grateful to MIT for its
hospitality and to Tokyo University for the extensive use of their
GRAPE systems.

\clearpage
\begin{deluxetable}{c|lc|lccc}
\footnotesize
\tablecaption{Core-collapse times for King models \label{tbl-1}}
\tablewidth{0pt}
\tablehead{
\colhead{} & \colhead{}   & 
\colhead{Isolated} & \colhead{}  & \colhead{} & \colhead{Tidally Truncated}   & 
\colhead{} \nl
\colhead{$W_0$} & \colhead{$t_{\rm cc}/t_{\rm rh}$}   & \colhead{$t_{\rm cc}/t_{\rm rh}$ (Quinlan)}   & 
\colhead{$t_{\rm cc}/t_{\rm rh}$}   & \colhead{$t_{\rm cc}/t_{\rm rh}$ (Quinlan)}   & 
\colhead{$M_{\rm final}$}  &  \colhead{$M_{\rm final}$ (Quinlan)}
} 
\startdata
1  & 18.1 & 17.89 & 10.0 &  5.98 & 0.30 & 0.10 \nl

2  & 17.9 & 17.85 & 10.8 &  7.74 & 0.37 & 0.17 \nl

3  & 17.7 & 17.61 & 12.0 &  9.49 & 0.44 & 0.24 \nl

4  & 17.3 & 17.24 & 12.9 & 11.26 & 0.53 & 0.33 \nl

5  & 15.9 & 16.37 & 13.3 & 12.73 & 0.64 & 0.44 \nl

6  & 13.9 & 14.49 & 12.4 & 12.94 & 0.76 & 0.57 \nl

7  & 10.6 & 10.84 & 9.30 &  10.50 & 0.86 & 0.72 \nl

8  & 5.32 & 5.79 &  5.21 &  5.76 & 0.88 & 0.85 \nl

9  & 2.10 & 2.25 &  2.01 &  2.25 & 0.96 & 0.92 \nl

10  & 0.86 & 0.93 & 0.80 & 0.93 & 0.97 & 0.96 \nl

11  & 0.41 & 0.47 & 0.40 & 0.47 & 0.99 & 0.98 \nl

12  & 0.20 & 0.26 & 0.20 & 0.26 & 0.99 & 0.99 \nl

\enddata
\tablenotetext{}{Core-collapse times for the sequence of isolated and
tidally truncated King models, computed using $N=10^5$ stars. Comparison is made with similar
results obtained by Quinlan (1996) using a 1-D Fokker-Planck method.}
 
\end{deluxetable}

\clearpage 
\begin{figure}[t]
\plotone{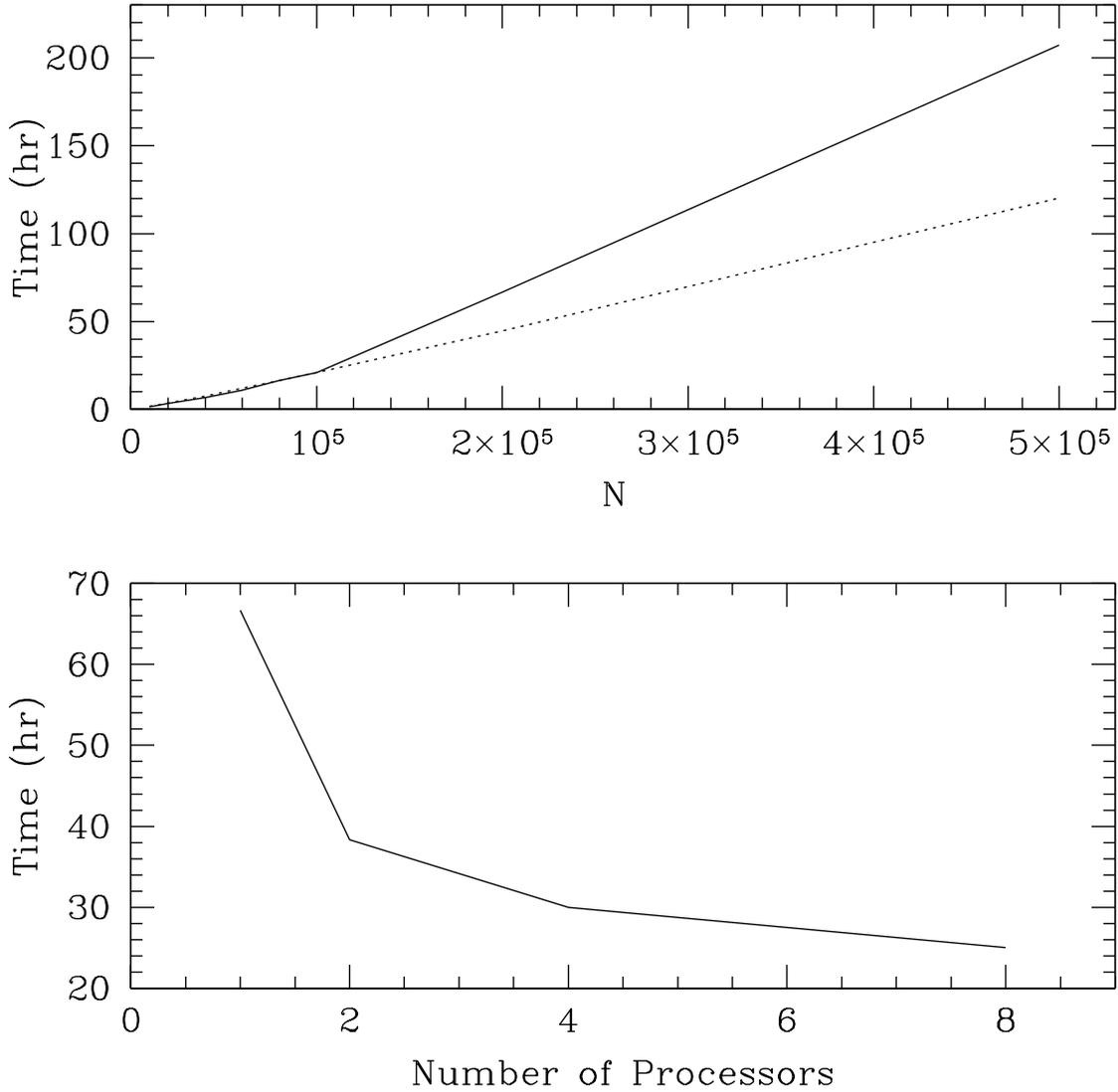}
\caption{The top frame shows the total computation time required (for an initial
Plummer model evolved up to core collapse) using one processor for simulations with up 
to $N = 5\times 10^5$. The dotted line indicates the theoretically estimated 
scaling of the computation time as $\sim N \log_2 N$. 
In practice, we find that the computation time scales as $\sim N^{1.4}$ for
$N = 1-5\times 10^5$. 
The bottom frame shows the scaling of the computation time (``wall-clock time'') 
with the number of processors for $N = 2\times 10^5$. 
\label{fig1}}
\end{figure}

\clearpage 
\setcounter{figure}{1}
\begin{figure}[t]
\plotone{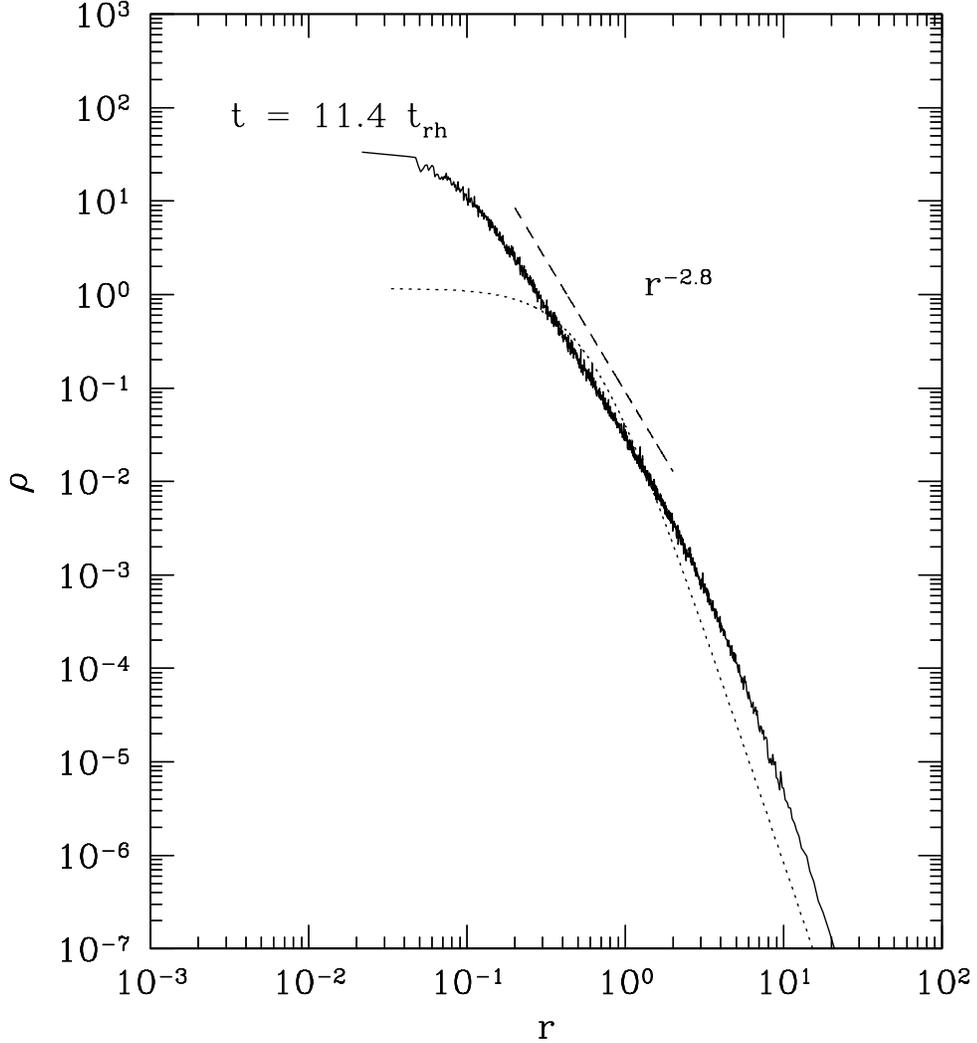}
\caption{(a) Density profile at an intermediate time, $t = 11.4\, t_{rh}$ during the 
evolution of an isolated Plummer model with $N = 10^5$ stars. 
The expected power-law in the density profile is clearly seen, 
with the best-fit exponent $\beta = 2.8$. The power law exponent approaches its
theoretical value of 2.2 as the cluster approaches core-collapse (cf. Fig.~2 b \& c).
The dotted line indicates the initial Plummer profile. 
Units are defined in \S 2.8.
\label{fig2a}}
\end{figure}

\clearpage 
\setcounter{figure}{1}
\begin{figure}[t]
\plotone{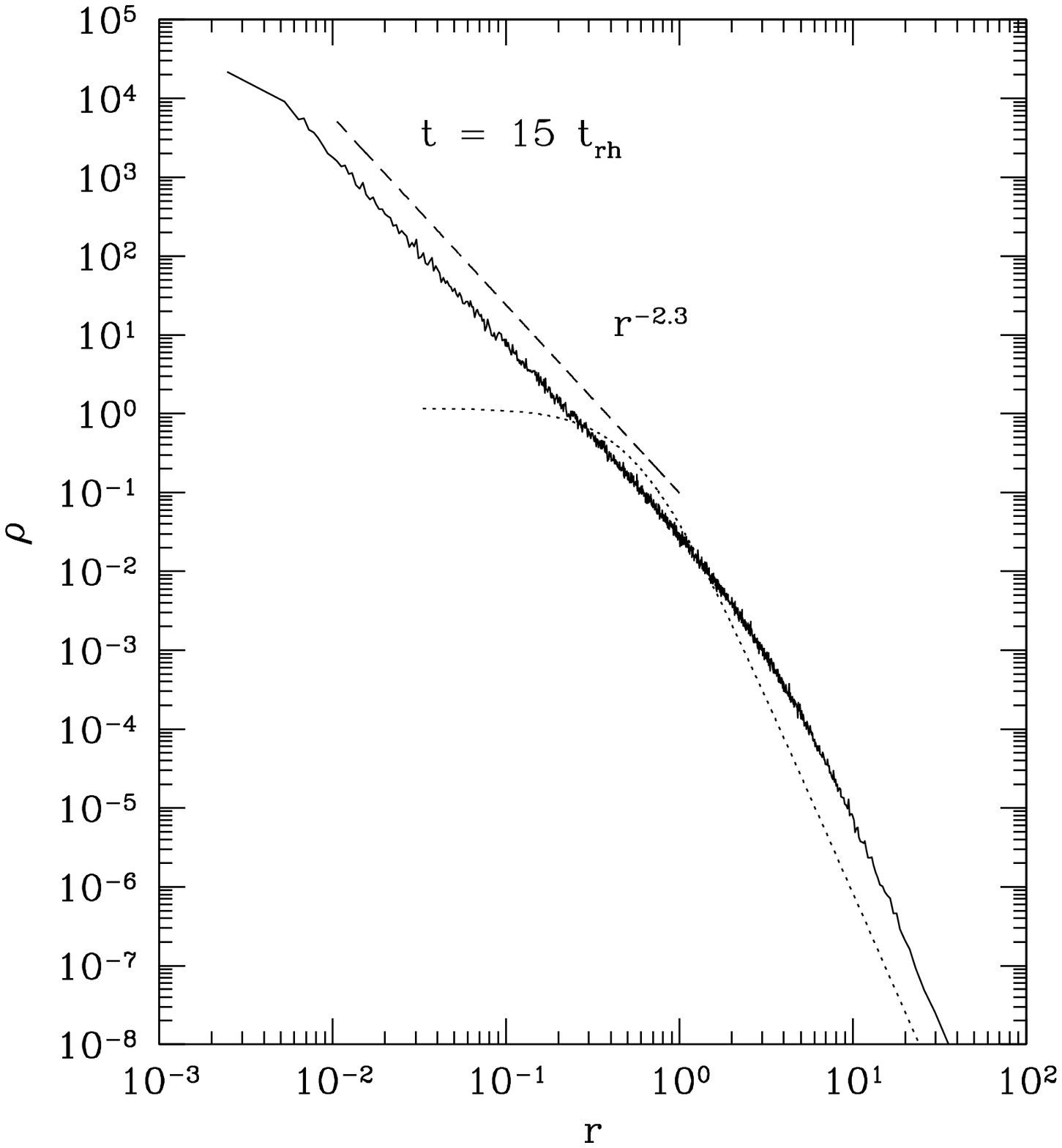}
\caption{(b) 
Density profile at $t = 15\, t_{rh}$ (just before core-collapse) for the
same model as in Fig.~2~a. 
The expected power-law in the density profile is now clearly seen, with the best-fit 
exponent $\beta = 2.3$, which is now closer to its theoretical value of 2.2.
The core density is about $10^4$ times greater than the initial density.
\label{fig2b}}
\end{figure}

\clearpage 
\setcounter{figure}{1}
\begin{figure}[t]
\plotone{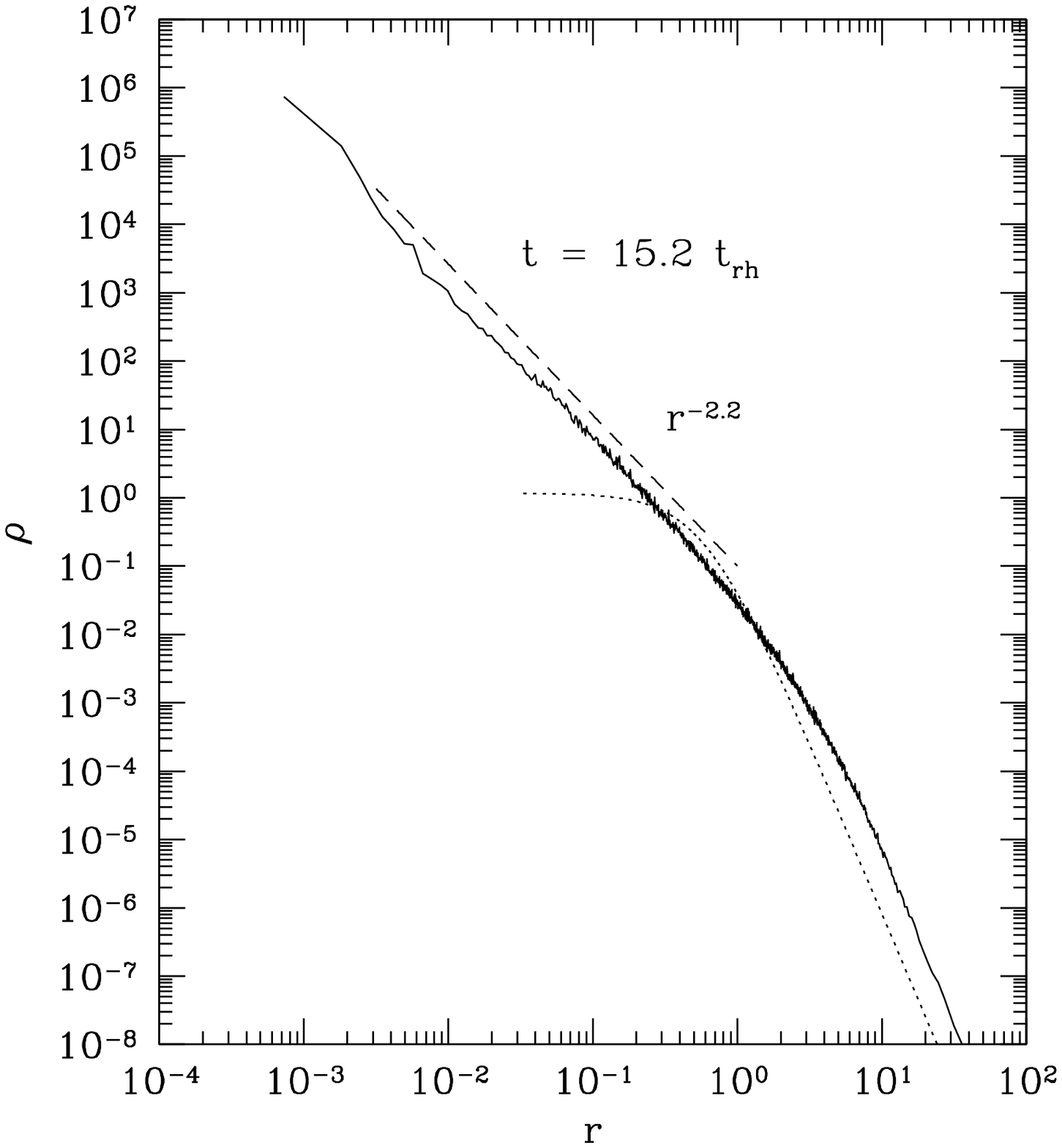}
\caption{(c) 
Density profile at $t_{cc} = 15.2\,t_{rh}$ (at core-collapse) for the
same model as in Fig.~2~a. 
The dashed line now indicates the \emph{theoretical} power law, with
exponent $\beta = 2.2$. The core density is almost $10^6$ times greater than
the initial density.
\label{fig2c}}
\end{figure}

\clearpage
\begin{figure}[t]
\plotone{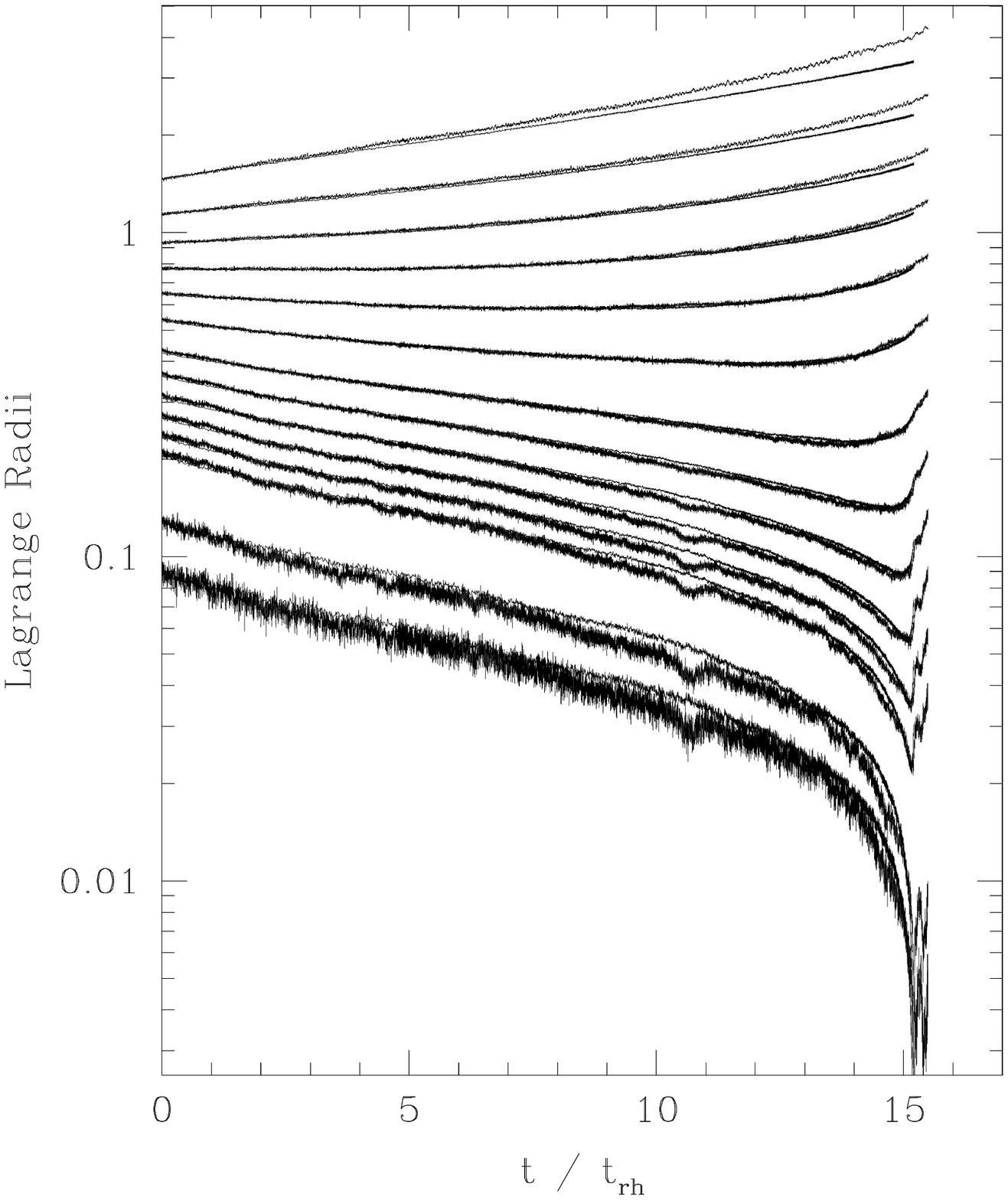}
\caption{Lagrange radii indicating the evolution of the Plummer model, 
with N = $10^5$ stars, compared with an $N$-body simulation with 
$N = 16384$ stars. Lagrange radii shown correspond to radii containing 
0.35, 1, 3.5, 5, 7, 10, 14, 20, 30, 40, 50, 60, 70, and 80 percent of the 
total mass. The Monte-Carlo simulation is terminated at core-collapse, 
while the $N$-body simulation continues beyond core-collapse. 
\label{fig3}}
\end{figure}

\clearpage
\begin{figure}[t]
\plotone{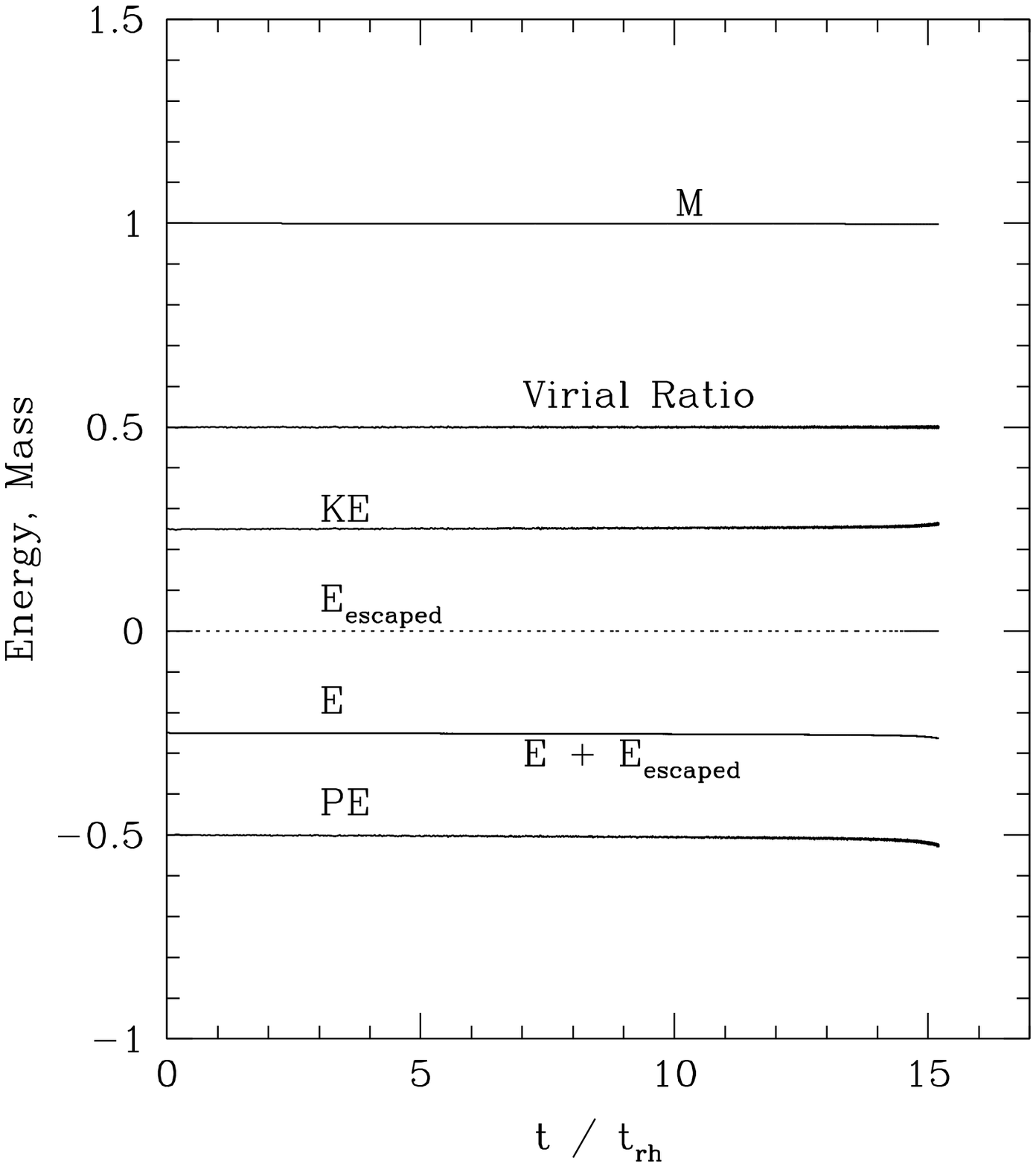}
\caption{The evolution of the total mass and energies for the same Plummer model as
in Fig.~3. The total mass loss at the time of core-collapse is $0.3\%$, and the
total energy loss is about 4\%. Most of the energy is lost during the late stages of 
evolution, with the energy loss up to $t = 10\,t_{rh}$ being less than 1\%. Here 
the energy carried away by escaping stars ($E_{escaped}$) is negligible.
\label{fig4}}
\end{figure}

\clearpage
\begin{figure}[t]
\plotone{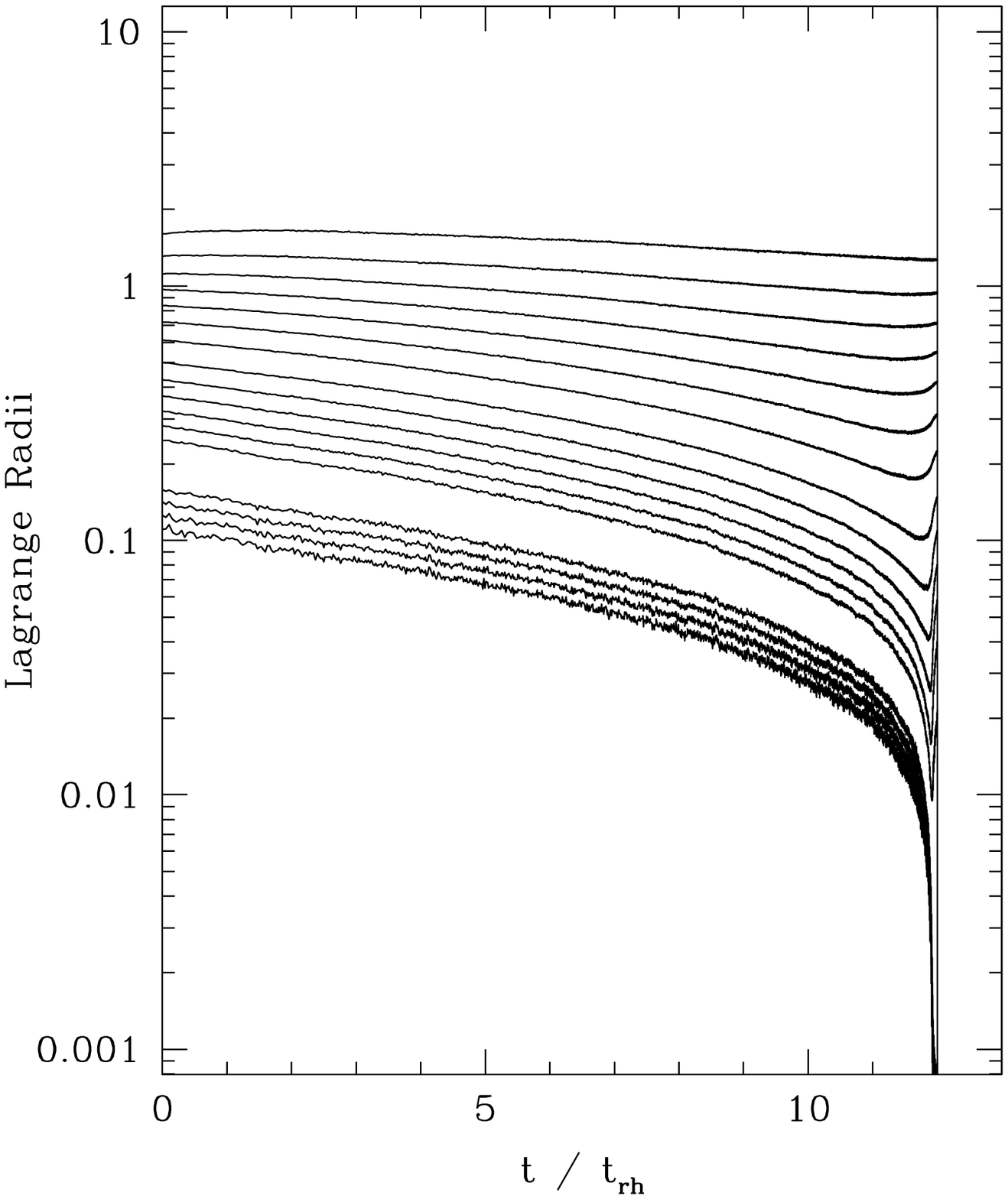}
\caption{Lagrange radii for the evolution of a \emph{tidally truncated} King 
model with $W_0 = 3$.
The tidal boundary causes stars to be lost at a much higher rate compared 
to the isolated model. The vertical line indicates the core-collapse 
time $t_{cc} = 12.0\,t_{rh}$. 
The presence of the tidal boundary reduces the core-collapse time by about 
32\% compared to the isolated model. In contrast, the evolution of an 
\emph{isolated} $W_0 = 3$ King model is very much like that of the Plummer
model shown in Fig.~3, with a total mass loss $< 1$\%, and 
$t_{cc} = 17.7\,t_{rh}$.
\label{fig5}}
\end{figure}

\clearpage
\begin{figure}[t]
\plotone{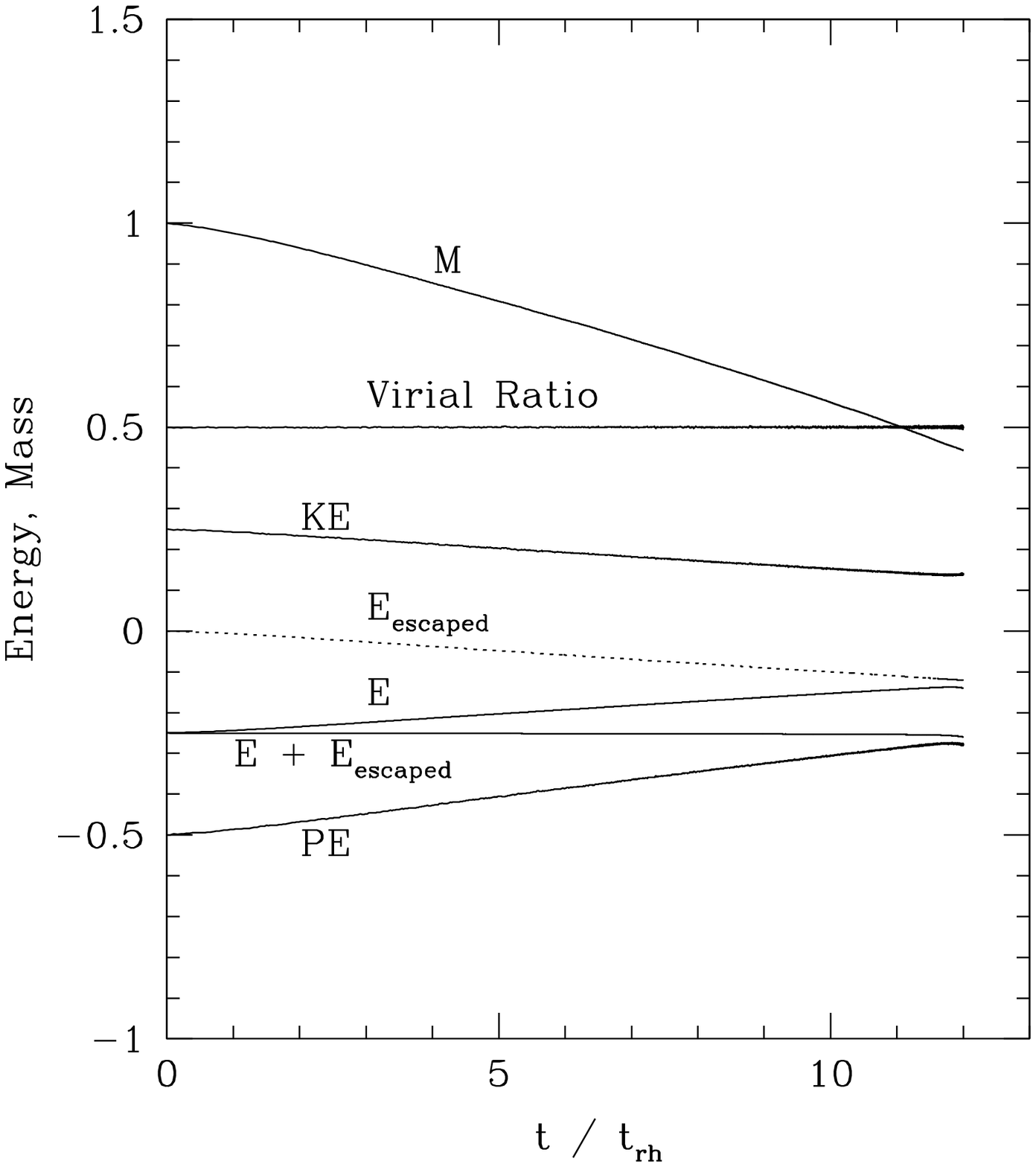}
\caption{The evolution of the total mass and energies for 
the model shown in Fig.~5. Only 44\% of the initial mass remains in the cluster
at core collapse. The dotted line indicates the energy carried away by escaping stars.
The large mass loss due to the tidal boundary causes the overall binding energy of the 
cluster to decrease significantly.
\label{fig6}}
\end{figure}

\end{document}